\documentclass[fleqn,usenatbib]{mnras}

\usepackage{newtxtext,newtxmath}

\usepackage[T1]{fontenc}
\usepackage{ae,aecompl}

\usepackage{graphicx}	
\usepackage{amsmath}	
\usepackage{nccmath}
\usepackage{siunitx}
\usepackage{multicol}
\usepackage{enumitem}

\newcommand{\be}{\begin{equation}}
\newcommand{\ee}{\end{equation}}
\newcommand{\bea}{\begin{eqnarray}}
\newcommand{\eea}{\end{eqnarray}}

\newcommand{\ifm}[1]{\relax\ifmmode#1\else$\mathsurround=0pt #1$\fi}
\newcommand{\kms}{\ifmmode\,{\rm km}\,{\rm s}^{-1}\else km$\,$s$^{-1}$\fi}

\newcommand{\ltsima}{$\; \buildrel < \over \sim \;$}
\newcommand{\lsim}{\lower.5ex\hbox{\ltsima}}
\newcommand{\gtsima}{$\; \buildrel > \over \sim \;$}
\newcommand{\gsim}{\lower.5ex\hbox{\gtsima}}

\def\la{\langle}

\def\M*{M_{\rm *}}

\def\Pi{\varpi_{_{\rm I}}}

\allowdisplaybreaks

\title[Distinguishing Between PI and CI Gas in the Circumgalactic Medium]{Distinguishing Between Photoionized and Collisionally Ionized Gas in the Circumgalactic Medium}
\author[C. Strawn et al.]{Clayton Strawn$^{1}$\thanks{E-mail: cjstrawn@ucsc.edu}, Santi Roca F\`abrega$^{2,3}$, Joel Primack$^{1}$\\
\smallskip 
$^1$ Physics Department, University of California, Santa Cruz, CA 95064, USA\\
$^2$ Universidad Complutense de Madrid, Departamento de F\'isica de la Tierra y Astrof\'isica, Madrid 28040, Spain\\
$^3$ Universidad Nacional Aut\'onoma de M\'exico, Instituto de Astronom\'ia de Ensenada, BC 22860, Mexico\\
}
\date{Accepted XXX. Received YYY; in original form ZZZ}

\pubyear{2022}

\begin{document}

\defcitealias{haardt_radiative_2012}{HM12}
\defcitealias{stern_does_2018}{S18}
\defcitealias{roca-fabrega_cgm_2019}{RF19}
\defcitealias{mandelker_instability_2020}{M20a}
\defcitealias{mandelker_lyalpha_2020}{M20b}

\label{firstpage}
\pagerange{\pageref{firstpage}--\pageref{lastpage}}
\maketitle

\begin{abstract}
\smallskip 
Most studies of highly ionized plasmas have historically assumed ions are either in photoionization equilibrium, PIE, or collisional ionization equilibrium, CIE, sometimes including multiple phases with different relevant mechanisms. Simulation analysis packages, on the other hand, tend to use precomputed ion fraction tables which include both mechanisms, among others. Focusing on the low-density, high temperature phase space likely to be most relevant in the circumgalactic medium, in this work we show that most ions can be classified as `PI' or `CI' on an ion-by-ion basis. This means that for a cloud at a particular point in phase space, some ions will be created primarily by PI and others by CI, with other mechanisms playing only very minor roles. Specifically, we show that ions are generally CI if the thermal energy per particle is greater than $\sim6$\% of their ionization energy, and PI otherwise. We analyse the accuracy of this ansatz compared to usual PIE/CIE calculations, and show the surprisingly minor dependence of this conclusion on redshift and ionizing background.
\end{abstract}

\begin{keywords}
Galaxies:Haloes -- Quasars:Absorption Lines -- Software:Simulations
\end{keywords}



\section{Introduction}
\label{sec:introduction}

The Circumgalactic Medium (CGM), the region of gas surrounding galaxies within their virial radius $R_{\rm vir}$ remains an enigmatic puzzle for the study of the evolution and development of galaxies and their environment. The existence of significant amounts of gas in this region, and its significance for star formation and structure formation has been well known for several decades. We have seen that galaxies, on their own, do not contain most of the baryons in the standard $\Lambda \rm{CDM}$ cosmology, and this problem is known as the missing baryon problem \citep[see][and references therein]{tumlinson_circumgalactic_2017}. Observations suggest that a significant fraction of the cosmic baryon budget is in the CGM of galactic haloes \citep{werk_cos-halos_2013,werk_cos-halos_2014}. The CGM also contains a large fraction, or perhaps even a majority, of the metals created in the Universe, as only about 20 -- 25 percent of the created metals remain in the galaxy in the form of stars, dust, and ISM gas. The mechanisms by which gas flows into and out of galaxies, while understood through broad-strokes `bathtub' models, have many details that are difficult to fully work out, such as the interaction between cool, inflowing streams,and hot, metal-rich outflows driven by supernovas or active galactic nuclei, as well as the effect of the interplay of this interaction with the contribution from virial shock heating \citep{birnboim_virial_2003,DekelBirnboim2008}. Also uncertain are the effects of magnetic fields \citep{DylanNelson2021}, cosmic rays \citep{hopkins_cosmic-ray_2020}, and thermal instabilities \citep{mandelker_instability_2020,mandelker_lyalpha_2020}. 

Due to its low density and high degree of ionization, it is very difficult to observe the CGM through emission except in very nearby galaxies or the Milky Way \citep{gupta_huge_2012,fang_xmm-newton_2015,AMIGA2020}. Instead, the primary way by which we study the CGM in the modern era is through absorption line spectroscopy. Background objects, mainly quasars, are used as light sources and their spectra are analysed to identify absorption lines and therefore detect what kind of gas is in the intervening clouds. The increased signal to noise of this kind of data, especially in recent years with the deployment of the Cosmic Origins Spectrograph (COS) on Hubble Space Telescope \citep{tumlinson_large_2011,werk_cos-halos_2013,werk_cos-halos_2014}, can give a very sensitive picture of a multi-component cloud of different ions. However the relatively small number of bright quasars means they rarely give multiple glimpses into the same galaxy, though there have been a few examples of multiple-detections either coincidentally \citep{lehner_evidence_2015,bowen_structure_2016} or using strong gravitational lensing to probe the same quasar in multiple places \citep{lopez_clumpy_2018,okoshi_multiple_2019}. With the successful launch of the James Webb Space Telescope in 2022, the new instruments should be able to acquire much better studies of the CGM, both in absorption and emission using the new MOS and IFU instruments \citep{kutyrev_microshutter_2008}.

To extract maximally useful information from the observations we do have, we often try to fit each velocity component of the gas as a separate phase. However, line confusion, saturation, and noise limits sometimes mean that there is some ambiguity about the possible mechanisms that can come into play. At CGM densities (which we will take to mean $10^{-8}$cm$^{-3} < n < 10^{2.5}$cm$^{-3}$), the primary mechanisms for ionizing a particle are photoionization and collisional ionization. In this low-density regime, recombination is effectively dominated by radiative recombination, and therefore only the photoionization and collisional ionization rates change with position in phase space. However, at the low temperature, high-density end, heavier elements can start to see the effects of other mechanisms, as detailed in Appendix \ref{sec:appendix}.

Photoionization, or PI, is where atoms absorb incoming photons from a variety of sources, including the metagalactic background radiation, emission lines from gas clouds, and radiation from stellar, AGN, and supernova sources. This state, assuming ions come to equilibrium, is called photoionization equilibrium, or PIE. In the CGM, outside of the region very near the galaxy \citep{sternberg_atomic_2002,sanderbeck_sources_2018}, the most important source is the metagalactic UV background. Generally, the most common background used by the community is that of \cite{haardt_radiative_2012}, hereafter \citetalias{haardt_radiative_2012}, and we also use that here unless otherwise specified. However, other UV background models have also gained attention in recent years \citep[e.g.,][]{faucher-giguere_new_2009,kuhlen_concordance_2012,Faucher2020}. 

In the PI case, the ionization level is almost a pure function of density with minimal temperature dependence. Effectively, denser clouds have fewer ionizing photons per particle, thus stabilizing with a higher fraction of low ionization states compared to high ionization states. So, using PIE, fitting the detected ions in a given component gives a good estimate of the density of the gas, which can be combined with the hydrogen column density to get an estimate of the metallicity, while the absorption line widths can give an estimate for the temperature.

Alternatively, atoms can be ionized through collisional ionization, or CI. When they collide, some of their kinetic energy is transferred to their internal electron structure, giving the electron(s) enough energy to escape. If only this mechanism is relevant, it is called collisional ionization equilibrium, or CIE. In this case, the ionization level is a pure function of temperature, and at higher temperatures, a greater proportion of gas is in high ionization states versus low ones. In CIE, fitting the detected ions gives an estimate of the temperature of the cloud, which when combined with the equivalent widths of the lines and relative amounts of different metal and hydrogen species, can give a good sense of the overall phase of the gas. 

There is much debate over which mechanisms are relevant and for which clouds of gas, and many different assumptions have been made to account for one, or the other, or both. At a superficial level, this difficulty is exacerbated by the fact that either mechanism taken by itself will clearly lead to pressure-balanced states in the CGM, which range from low-density hot gas to high-density cool gas, to contain higher and lower ions, respectively. However, to analyse observations in a sophisticated way by including both mechanisms can be very difficult, due to having more difficulty efficiently constraining either density or temperature. If the mechanism cannot be assumed, then both variables need to be decided by the noisy properties of the lines themselves, and the appearance and relative strength of different ions cannot be used to constrain either quantity. 

In modern galaxy simulations, by contrast, determining the phase of gas in the simulated CGM is not itself a challenge, though correctly evolving the phase remains quite unsolved, with different codes leading to vastly different results even with the same initial conditions (e.g. the {\sc agora} project, \citealp{roca-fabrega_agora_2021}, Roca-F\`abrega et al. in prep., Strawn et al. in prep.). Since in a simulation the full physical state of every parcel of gas is easily available, there is no need to assume only one mechanism is relevant. Software codes like {\sc trident} \citep{hummels_trident_2016}, {\sc pygad} \citep{rottgers_absorption_2020}, and others \citep[e.g.][]{churchill_ionization_2014,churchill_direct_2015} simply interpolate pre-made tables from {\sc Cloudy} \citep{ferland_cloudy_1998,ferland_2013_2013,ferland_2017_2017} to determine ionization fractions as a function of both temperature and density, without any need to explicitly reference the two mechanisms. However, we believe there is still some value in defining certain ions to be created predominantly through PI or CI, incorporating some of breakthroughs in simulation studies. In particular in \cite{roca-fabrega_cgm_2019} and \cite{strawn_o_2021}, we used a definition of PI-dominated and CI-dominated gas to distinguish O~{\sc vi} states in a simulation. This definition led to the discovery within the cosmological simulation of a thin CI-O~{\sc vi} interface layer on the edge of cool, inflowing PI-O~{\sc vi} clouds. 

This definition of PI and CI leads to an ion-by-ion distinction where some ions are predominantly determined by PI mechanisms, and others are predominantly determined by CI mechanisms. By splitting up these two types of ions, the weakness of using a full PI and CI model (requiring both temperature and density of a cloud be determined by the absorption line shapes) can be mostly alleviated, as one can use a rough temperature to determine which mechanism is most relevant for a component, and then constrain the density with PI ions, and/or the temperature with CI ions.

This paper is organized as follows. In Section \ref{sec:definition} we discuss the definitions of PI and CI gas for different ions, and show which ions are in which state as a function of density and temperature throughout phase space. In Section \ref{sec:redshift-dependence}, we analyse the effect of changing the extragalactic background, either with \citetalias{haardt_radiative_2012} but at different redshifts or by arbitrarily modifying the central powerlaw of \citetalias{haardt_radiative_2012} according to the procedure outlined in \cite{haislmaier_cos_2021}, to show that none of these changes meaningfully affect the distinctions used here. In Section \ref{sec:ionization-energy}, we discuss the physical principle at play here, and show that for each ion, where the cutoff between CI and PI occurs depends on the ratio of its ionization energy to the average energy per particle at that temperature. In Section \ref{sec:split-model} we outline the effect of using PIE for some regions of phase space, and CIE for others, depending on the detected ions. Finally, we summarize our conclusions in Section \ref{sec:conclusion}. 

\section{Ion-by-ion Definition of CI and PI}\label{sec:definition}

We generate all the data necessary for this definition from the software {\sc Cloudy}\footnote{We use version {\sc Cloudy 17.03} for the data in this work, see \cite{ferland_2017_2017}}. {\sc Cloudy} is an open-source spectral synthesis code which simulates the state of gas under many different astronomically-relevant physical conditions. It is at the basis of many, if not most, gas physics packages used in modern simulation codes. Among its many other uses in the community, of particular interest to us is that it tracks the distribution of each element into different ions as a function of density, temperature, time, and location relative to different ionizing sources, and intervening absorbers. 

The way {\sc Cloudy} is used in simulations of the circumgalactic medium, as in {\sc trident}, {\sc pygad} and all their dependent papers, e.g. \citet[][]{peeples_figuring_2019,strawn_o_2021} and \citet[][]{rottgers_absorption_2020}, respectively, is to always make a few simplifying assumptions that allow a small number of pregenerated tables to define the ion fractions in each cell (or, in each smoothed gas particle) of the simulation. There are two common assumptions that go into this simulation. 

First, it is assumed that the different ions are always in ionization equilibrium, where the rate of particles entering a particular ionization species through ionization of lower states and through recombination from higher states is equal to the rate of particles leaving that state through further ionization or recombination into lower states. In other words, it is assumed that ionization equilibration, in all areas of phase space, takes place much faster than changes in temperature or density. This approximation is generally fairly good \citep[see, for example,][]{ji_simulations_2019} as long as there are not too extreme of energy events, such as near AGN, that can leave some `fossilized' ionization for thousands of years directly in the path of an AGN jet, even long after the AGN has shut off \citep{oppenheimer_agn_2013}. In any case, analysis of simulations through post-processing cannot be done without this assumption, or an equivalent non-equilibrium ion fraction table, which is a function of recent temperature and/or density. In principle ion fractions could be tracked over time as separate fields within the simulation, and thus evolve from the ionization and recombination rates directly, but this calculation would greatly increase computational time and expense, and require significantly finer timesteps than are possible today.

Second, it is assumed that ionizing radiation is uniform and isentropic. Uniformity is enforced by inserting a constant ionizing background radiation spectra, which does not depend on any local sources or effects. The most common background used is \citetalias{haardt_radiative_2012}, though other possible backgrounds are also relevant. When relatively close to a galaxy, especially if it is undergoing a starburst or AGN activity, this approximation is not very good \citep{sternberg_atomic_2002,sanderbeck_sources_2018}. However, an improved schema which is not yet in widespread use but may be soon would be to use some precomputed {\sc Cloudy} tables with different, realistic backgrounds according to the approximate distance to the galaxy centre \citep[e.g.,][]{Gnedin2012,Kannan2014,Kannan2016}. 

Isentropy is essentially a claim that the CGM is optically thin, and therefore the direction from which a photon comes has no impact on its penetration into the material. Gas in the CGM is usually fairly low density (typical number densities are between $10^{-7}$ and $10^{-1}$ cm$^{-3}$), and so this assumption should have little impact on our results. However, there are other regions of the CGM which might be higher density and therefore self-shielded \citep{omeara_hst_2013,altay_through_2011}. While this effect is generally accounted for directly in subgrid models for heating and cooling \citep[see, e.g.][]{kim_agora_2016}, it is not accounted for by our ion fraction grid.

\begin{figure}
\includegraphics[clip, width=0.99\linewidth]{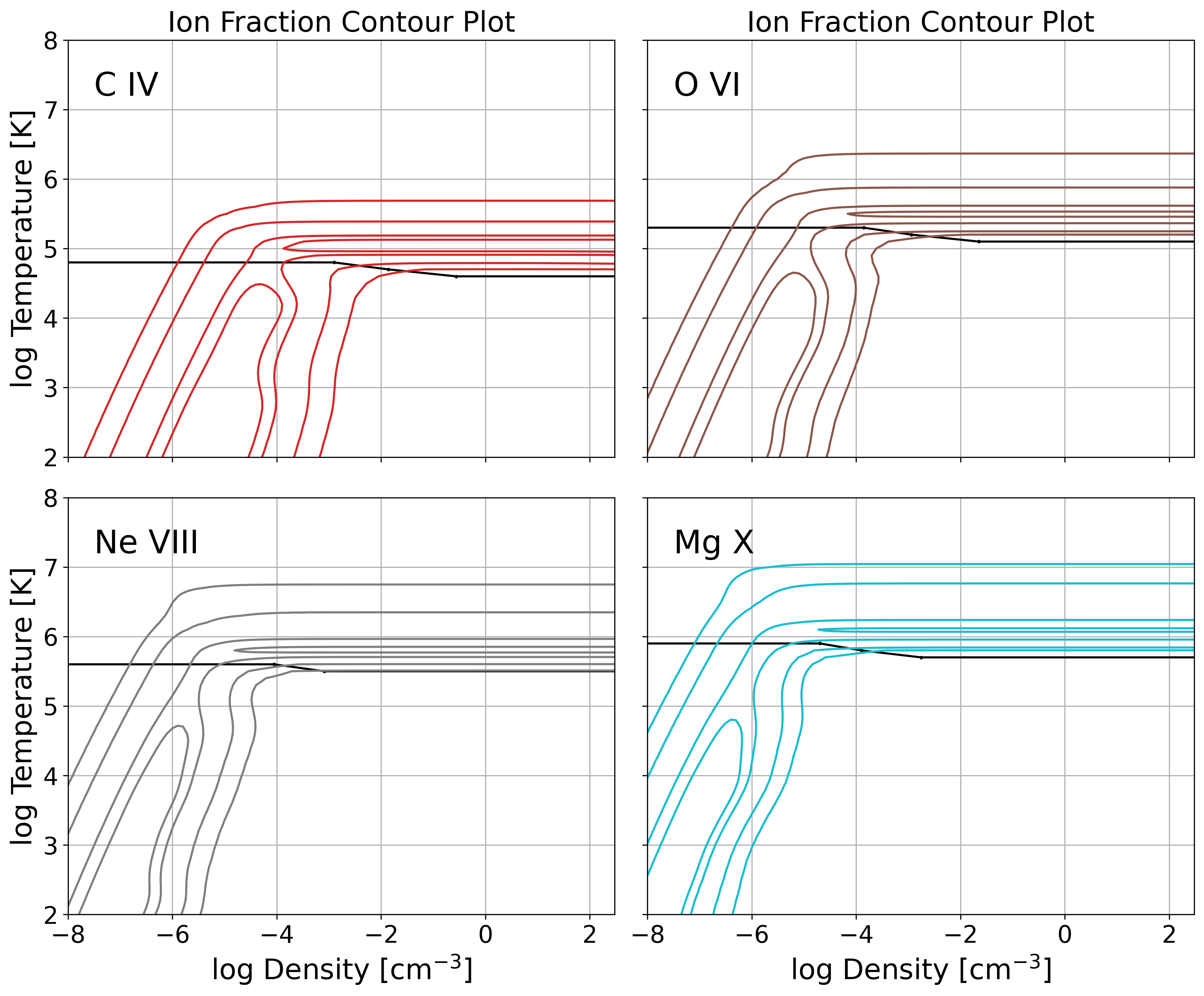}\par 
\caption{Contour plots for C~{\sc iv}, O~{\sc vi}, Ne~{\sc viii}, and Mg~{\sc x}. Each has a CI peak (upper right) and a PI peak (lower left). The definition in Section \ref{sec:definition} distinguishes the two via the black line: CI above and PI below.}
\label{fig:contours}
\end{figure}

With these assumptions made, {\sc Cloudy} can create accurate tables of ion fraction as a function of temperature, density, and ionizing background. A contour plot of these fractions with the background from \citetalias{haardt_radiative_2012} at redshift $z=0$ is shown in Fig. \ref{fig:contours}. We focus here on some of the species that are particularly well-studied in observations and simulations, because they have very strong lines due to their lithium-like nature (3 electrons remaining), although their significant differences in charge means they have a broad range of ionizing potentials. Besides these lithium-like ions, the most commonly studied other ions in observation are low ions (neutral, singly or doubly ionized). These ions are often not studied in cosmological simulations due to the expected cloud sizes being too small to effectively resolve (\citealp{peeples_figuring_2019}, \citealp{hummels_impact_2019}; but see also \citealp{DylanNelson2021}).

Each contour plot can easily be broken up into a PI region, on the lower left side, and a CI region, on the upper right side. The PI gas is weakly dependent on temperature and strongly dependent on density, while the CI region, is totally independent of density (above some critical density), and depends strongly on temperature. 

While the two peak regions are easily identified by eye, researchers in the CGM are still interested in giving a rigorous definition that correctly categorizes the mechanisms in regions far from both peaks. Doing so has proven somewhat difficult, despite many attempts over the last few years. Among others, these include separating the regions by density \citep{faerman_massive_2020}, analyzing the gas PI and CI timescales \citep{churchill_direct_2015}, separating by temperature \citep{sanchez_not_2019}, and artificially restricting to only two obvious options and then making a binary judgement \citep{stern_does_2018}. 

\begin{figure}
\includegraphics[clip,trim={0.0cm 0.0cm 0.0cm 0.0cm}, width=0.99\linewidth]{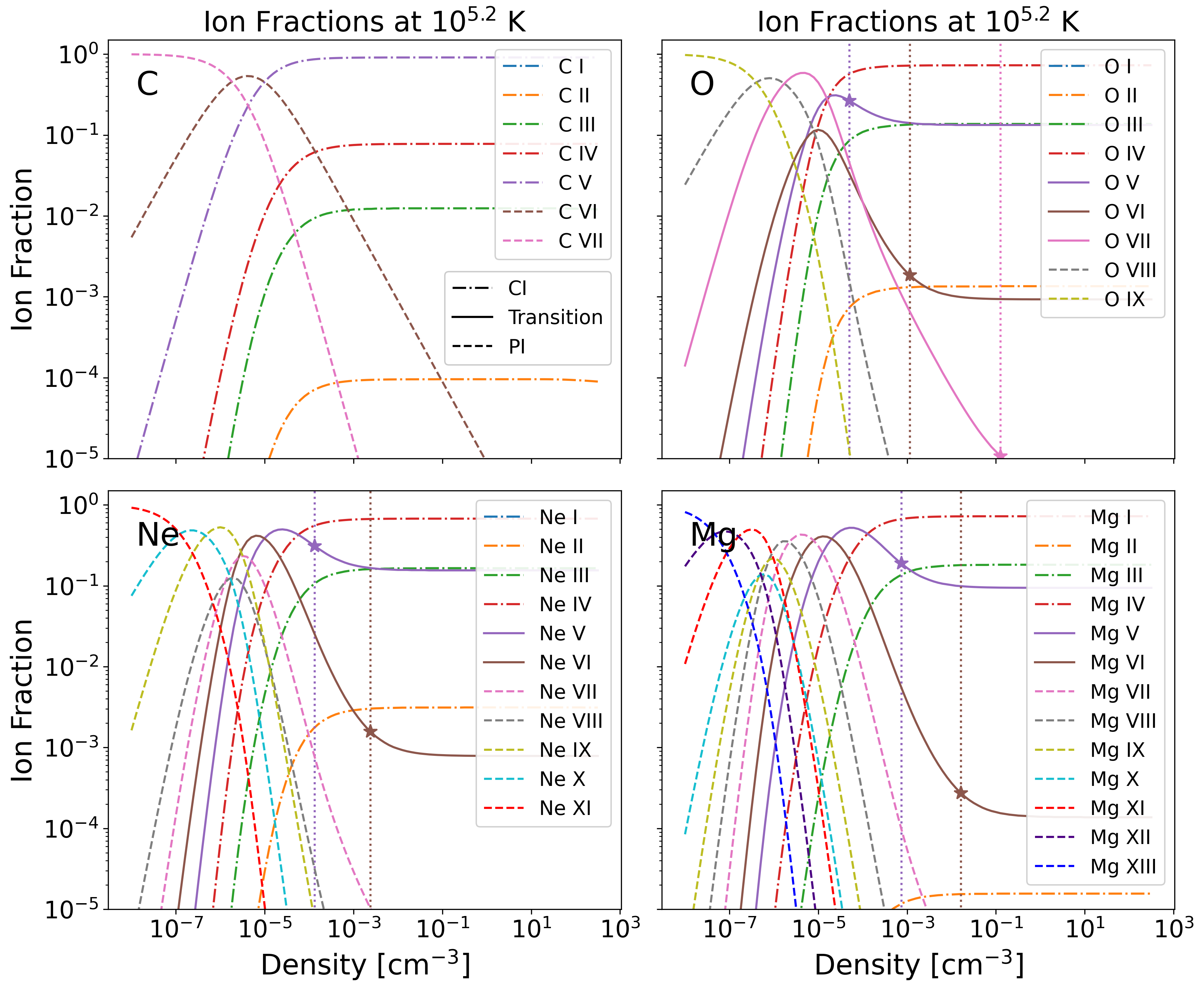}\par 
\caption{Ion fraction vs density for all species of four atoms: Carbon, Neon, Oxygen, and Magnesium. All four images are at a fixed temperature of $T = 10^{5.5}$ K. The linestyle reflects what mechanism the algorithm in Section \ref{sec:definition} identifies as dominant, with dashed indicating PI, dot-dashed indicating CI, and solid indicating transitionary. For transitionary ions, the transition density is also shown with a vertical dotted line and star of the same color. Negligible species at this temperature (i.e., neutral) are not plotted.}
\label{fig:fractions}
\end{figure}

In \cite{strawn_o_2021}, we showed, based on the work of \cite{roca-fabrega_cgm_2019}, that there is a straightforward way to define the contributions of the two phases based on a physical argument, rather than purely on these contour plots. This argument allows the definition to clearly extend to regions far from both peaks, such as in the transition region between them or the high-density, low-temperature corner. The process begins with fixed-temperature fraction-density curves, as seen in Fig. \ref{fig:fractions}. The density dependence can be interpreted as follows: photon density per particle increases as density decreases, here meaning when tracking these graphs from right to left. Thus an increase of ion fraction when moving leftward indicates ions being created through photoionization, and a decrease indicates ions being destroyed through photoionization. 

There are three possible shapes of fraction-density graphs at fixed temperature at CGM densities. They are characterized by the existence, or nonexistence, of a maximum density, and by a flat shelf at high density, which we will refer to as the `CI base'. 

\begin{itemize}
    \item \textbf{`Collisionally Ionized'}: First, the graph can stabilize to a CI base at high density, and always decreases in fraction with decreasing density. This gas is called `Collisionally ionized' as PI processes only destroy, rather than create, this ion at this temperature.
    \item \textbf{`Photoionized'}: Second, they can fail to stabilize to a CI base at high density. Instead, with decreasing density, they first increase up to a maximum, and then decrease back to zero. This gas is called `Photoionized' as PI processes create all of this ion at this temperature.
    \item \textbf{`Transitionary'}: Third, they can have both a maximum and a CI base, meaning they stabilize at high density, but still increase from that value as density decreases. We call this gas `transitionary' and the `transition point' is defined as the density where the CI base is equal to 50 percent of the total (the rest coming from PI). At densities on the right of the increase, the ion is created mostly through CI, and on the left, the ion is created mostly through PI. 
\end{itemize}

\begin{figure*}
\includegraphics[clip,trim={0.0cm 0.0cm 0.0cm 0.0cm}, width=0.99\linewidth]{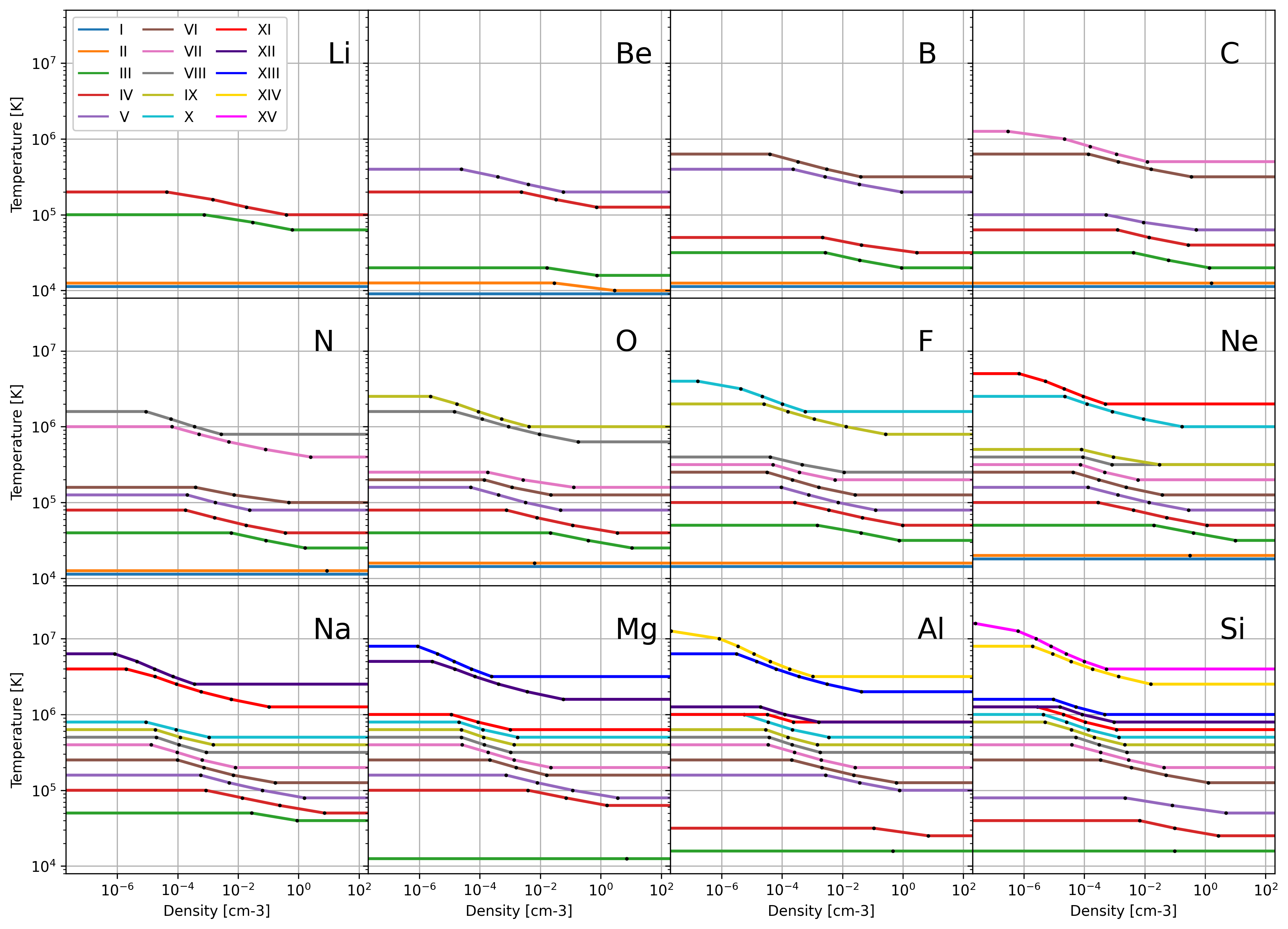}\par 
\caption{PI vs CI cutoffs for all species of the first 12 metals. Above each line, the indicated ion is CI, and below the line, it is PI. Some cutoffs are overlapping, and are shown with slight offsets to see multiple colors at once. Lower ions always have cutoffs at lower temperatures than high ions, but a nonsequential colormap is used to facilitate identifying ions of interest. Neutral and singly ionized states do not appear for the bottom row, see Section \ref{sec:neutral-single-ionized} and Appendix \ref{sec:appendix} for details.}
\label{fig:cutoffs}
\end{figure*}

The whole of phase space can thus be characterized for each ion (with some subtlety needed for neutral, singly-ionized, and fully ionized states, see Section \ref{sec:neutral-single-ionized}): at some temperatures, an ion can be characterized as fully created through PI, at other temperatures, it can be characterized as fully created through CI, and at still other temperatures, it is primarily PI below and CI above a particular density. In Fig. \ref{fig:cutoffs}, we show the full z=0 distribution for all species of the first 12 elements. As expected, every ion is PI at low temperatures, CI at high temperatures, and transitionary at intermediate temperatures. However, the change does not happen at the same temperature for all ions. Higher ionization states remain primarily PI at much higher temperatures than lower ionization states, so in a single cloud of gas, some low ions can be present that were created through CI while high ions are also present that were created through PI.
\subsection{Neutral, Singly-ionized, and Fully ionized states}\label{sec:neutral-single-ionized}

The above analysis can be expanded to capture the basic structure of atoms even where the ionization state is harder to define. For neutral atoms, there is first of all a semantic point. By definition, neutral atoms are not ionized, so in principle they are not `created' through PI or CI structures, rather it would be the absence of either. We will still use the terms `PI' and `CI' for neutral atoms, but by this we simply mean `as if the fraction was determined through PIE' or `as if the fraction was determined through CIE', respectively.

The more serious problem is that, given a specific temperature, at no density could there possibly be a maximum for the neutral fraction. This is because the neutral fraction at all temperatures should be a monotonically increasing function of density. A `PI' state for a neutral atom requires the highest density to be counted as a maximum, and a `Transition' state is impossible. So, we will define the difference between CI and PI to simply check for the existence of a CI base on the high-density end which remains sufficiently flat. 

The fact that there is no possibility of a transition temperature for neutral species also leads to complications for the first-ionized state (C II, O II, etc.), because all transitions out of the neutral state will directly enter the singly-ionized state, no matter which mechanism is used. In other words, the neutral atom can only have a CI base if the singly-ionized state also has a CI base at the same time. Thus, we add some additional considerations to the algorithm that if either the singly-ionized or neutral state is PI, the other will be also. For most atoms, this algorithm gives a transition for singly ionized states around T$=10^{4.1}$ K. 

However, neutral and singly ionized states for `heavy elements', here meaning elements in the third row of the periodic table (here Na, Mg, Al, and Si) are exceptions to this rule. In these cases new shapes not described in Section \ref{sec:definition} can appear, and therefore cannot be categorized as PI or CI. This effect arises because due to both their low ionization energy and their quantum structure, these ions have a much more tenuous hold on their outer electrons than the rest of the species often seen in the CGM. At low temperatures and high densities, their ionization mechanism does not conform to the classification algorithm, because a third mechanism, charge-exchange \citep{dopita_astrophysics_2003, kallman_photoionization_2021}, becomes comparable to photoionization. This difficulty reiterates the fact that this classification is only a simplification that applies at the low densities characteristic of the CGM and with ions that have relatively tightly bound electrons. The lowest ions for heavy elements therefore have no meaningful `PI dominated region', and cannot `transition' from PI to CI. As such, they will be excluded from analysis for the rest of the paper. See Appendix \ref{sec:appendix} for further details. Unfortunately, this means one of the most commonly studied ions in absorption, Mg~{\sc ii} \citep[e.g.][in simulations and observations, respectively]{DylanNelson2021, darekar_probing_2022}, cannot be simply classified as CI or PI.

Fully ionized states with no electrons remaining in principle also could require a more subtle definition. Symmetrically to the neutral state, there can be no maximum at any temperature, only a monotonic (now decreasing) function with increasing density. However by taking the lowest-density fraction as the maximum, we find this situation remains remarkably similar to the standard case, and no special treatment is needed. The algorithm detects a transition if a CI base exists, and, when moving from high density to low density, the ion fraction increases by a factor of two or more from the CI base to the maximum. The ion is CI at all densities if this change is never that large, and PI at all densities if there is no base to speak of at that temperature. We see in Fig. \ref{fig:cutoffs} that fully ionized states become created entirely through CI at between $T=10^{6}$ K, for C~{\sc vii}, and $T=10^{7}$ K, for Mg~{\sc xiii}. We will show in Section \ref{sec:ionization-energy} that this outcome aligns perfectly with the theoretical prediction of all other states, and thus we will not modify the algorithm to account for these ions as we did for neutral and singly ionized states.

\section{Redshift and Background Dependence}\label{sec:redshift-dependence}

One might anticipate that the above results are strongly dependent on ionizing background radiation. After all, \citetalias{haardt_radiative_2012} already shows an extremely wide range of ionizing intensities with redshift, with basically no background radiation at $z=10$, to a peak at $z=2$, and then winding back down to the value observed today in nearby galaxies. One might expect that a stronger ionizing background leads to a higher proportion of ions created through PI. But it turns out that the ionizing background has a surprisingly small effect on the conclusion about where in phase space an ion is created through primarily PI or CI mechanisms, at least over the redshift range $z=0-4$. The reason is primarily the fact that increasing the ionizing radiation, as long as it is at least somewhat uniform and doesn't have any outrageous spikes at particular frequencies, always increases the number of ions photoionized into a state at the same time as it increases the number further photoionized out of that state. At a higher overall level of background radiation, the overall ionization of the whole population of ionic species shifts to higher ionization, but each individual species is only slightly moved, and only in a very small region does it actually change the dominant mechanism. 

\begin{figure}
\includegraphics[clip,trim={0.0cm 0.0cm 0.0cm 0.0cm}, width=0.99\linewidth]{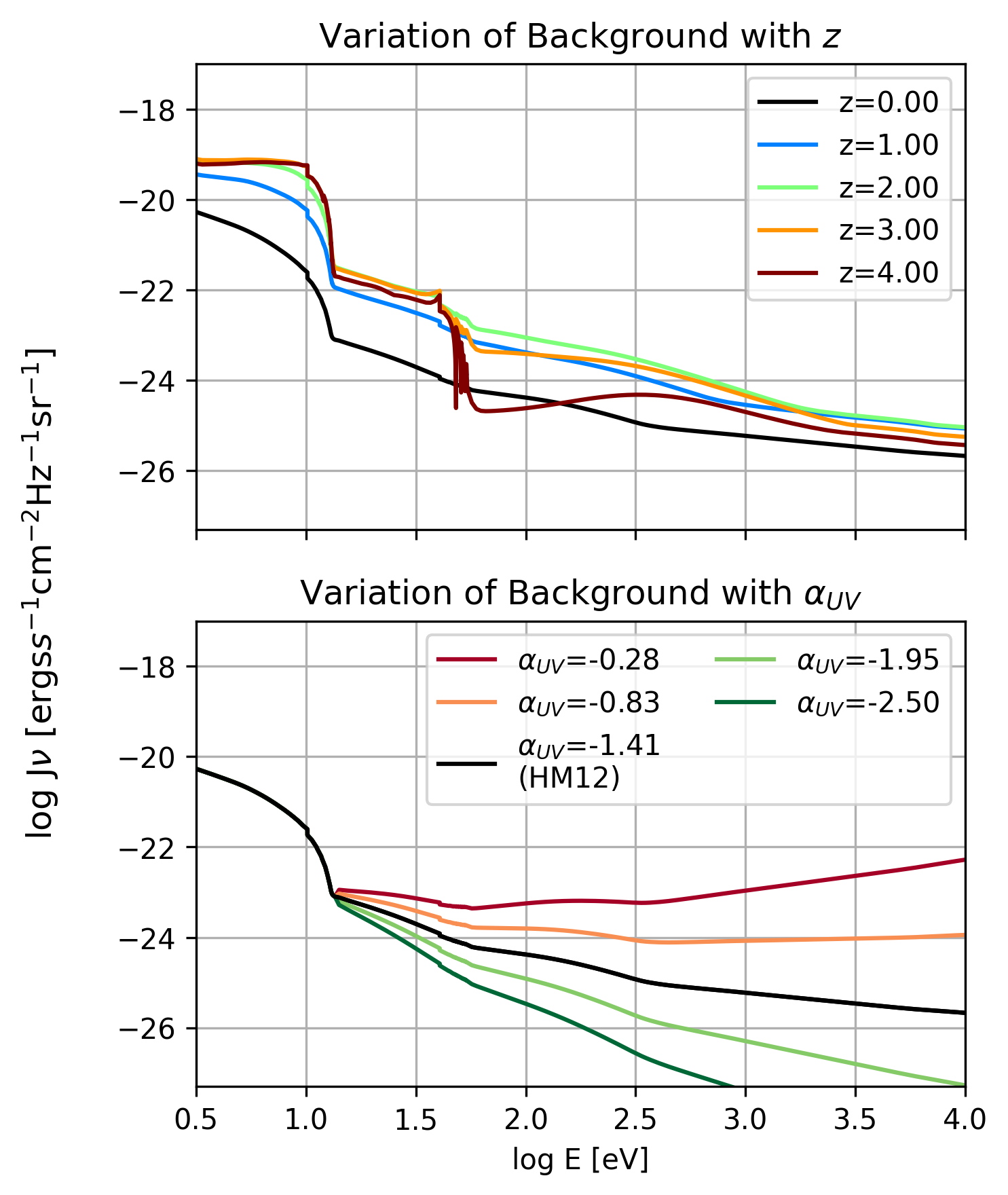}\par 
\caption{Top, evolution of the 
\citetalias{haardt_radiative_2012} UVB with changes in redshift from $z=4$ to $z=0$. Bottom, effect of changes in UVB by artificially varying the powerlaw slope $\alpha_{UV}$. In both panels, the black line is the fiducial \citetalias{haardt_radiative_2012} $z=0$ spectrum used throughout this paper except when otherwise specified.} 
\label{fig:uvb}
\end{figure}

We test this dependence in two ways, first by checking the results of this procedure with tables generated by {\sc Cloudy} with \citetalias{haardt_radiative_2012} at $z = 0$, $1$, $2$, $3$, and $4$, and then by checking the results by arbitrarily varying the slope of the ionizing background around the $z=0$ fiducial shape. The formula for this modification is taken from \citet{haislmaier_cos_2021}, and was first used in \citet{crighton_metal-enriched_2015}, \citep[see also][]{fumagalli_physical_2016}. At energies greater than 1 Rydberg, the power-law slope of \citetalias{haardt_radiative_2012} is approximately $\alpha_{UV} = -1.41$. We replace the \citetalias{haardt_radiative_2012} with a new background with $\log_{10} F_\nu = f(E)$, where $f(E)$ is defined as
\begin{ceqn}
\begin{equation}
\label{eq:alpha_variation}
  f(E) =
    \begin{cases}
      H(E), & E \leq E_0 \\
      H(E) + (\alpha_{UV}+1.41)\cdot\log_{10}(E/E_0), & E > E_0 \\
    \end{cases}       
\end{equation}
\end{ceqn}
Here $H(E)$ is the base 10 logarithm of \citetalias{haardt_radiative_2012} and $E_0$ is 1 Rydberg, or $13.6$ eV. In Fig. \ref{fig:uvb} we show the difference in ionizing background caused by varying these two quantities.

\begin{figure}
\includegraphics[clip,trim={0.0cm 0.0cm 0.0cm 0.0cm}, width=0.99\linewidth]{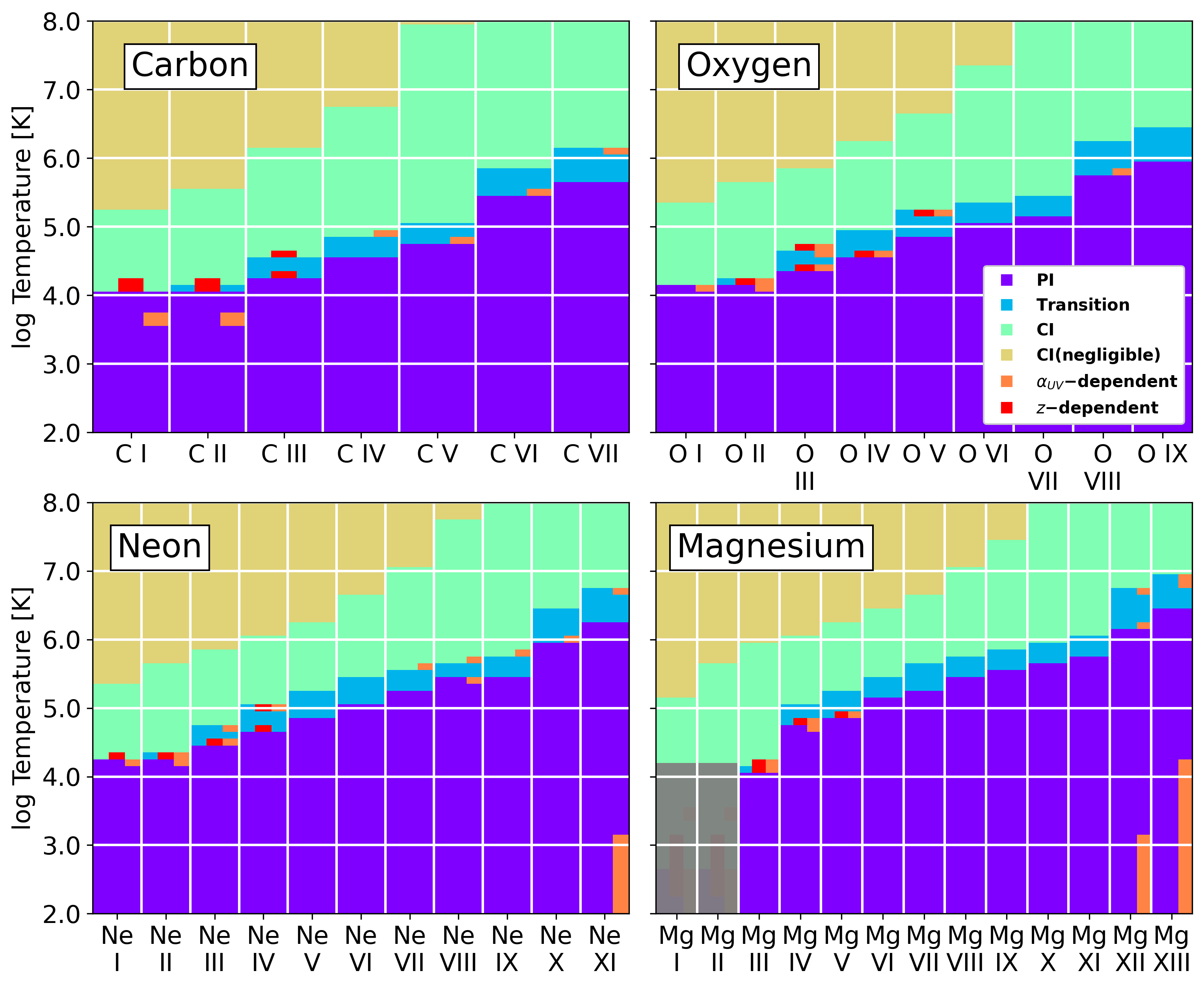}\par 
\caption{Ionization mechanism vs temperature for all species of Carbon, Oxygen, Neon, and Magnesium, computed on a grid of 0.1 dex in temperature space. The colors besides red and orange show the mechanism at redshift $z=0$ with the fiducial background. Red bars indicate grid points which had any change from the $z=0$ mechanism at $z = 1, 2, 3,$ or $4$., and orange bars indicate the same, but for any change from the \citetalias{haardt_radiative_2012} ($\alpha_{UV} = -1.41$) mechanism at $\alpha_{UV} = -0.28 , -0.835, -1.945,$ and $-2.5$. The grayed-out region is not classified properly with this scheme, see Appendix \ref{sec:appendix}. Even with significant changes in background, changes in mechanism are shown to be minimal.} 
\label{fig:temp-zdep}
\end{figure}

In Fig. \ref{fig:temp-zdep} we show the overall distribution of ionization mechanism with temperature at redshift $z=0$. All ions follow the same trend of being PI below some temperature, transitionary for a small number of temperature steps (sometimes zero), and then CI above. Of course, the CI fraction becomes negligible at high enough temperature, and those regions are indicated in gold. The red bars here indicate regions where the dominant mechanism is changed with changes in redshift from $z=4$ to $z=0$. Effectively, the change in the ionization mechanism's temperature dependence is negligible, with the differences confined mostly to a single data point per ion (spacing being 0.1 dex in temperature). Even this change generally only happens with low ions. Magnesium is a heavy element in our schema, so Mg~{\sc i} and Mg~{\sc ii} at low temperatures cannot be classified with either mechanism in this scheme (see Appendix \ref{sec:appendix}), which is why that region is greyed out. The orange bars indicate ions which have different mechanisms at the same temperatures with changes in $\alpha_{UV}$. Again, differences are fairly rare, however in this case they are more likely on high ions, presumably because the ionizing photons for highly ionized species are more affected by the changes to $\alpha_{UV}$. Affected species include, interestingly, the highest ions at the lowest temperatures. This has the physically intuitive meaning that with very soft backgrounds ($\alpha_{UV} = -1.945, -2.5$), there are not enough high-energy photons to photoionize all the way to Mg~{\sc xiii} or Ne~{\sc xi}, and the PI `maximum' the algorithm searches for is never detectable. It therefore classifies those states as CI at low temperatures, because they are not PI-accessible with this extremely soft background.

We remind the reader however that this redshift independence does not mean that there will be no evolution in the relative significance of the two mechanisms with redshift or background when considering an individual ion. In \citet{roca-fabrega_cgm_2019} and \citet{strawn_o_2021} we showed that changes in redshift accompanied changes and even reversals in PI or CI dominance for O~{\sc vi}. The difference here is that we are not analyzing a cosmological simulation, but a simple grid in density-temperature space. Gas processes like heating, cooling, and chemical evolution cause metals in the CGM to occupy different regions of this graph with time, and thus affect the `dominance' of one mechanism in the region as a whole. In fact, the conclusions here suggest that the change in PI-CI ratio over time is almost entirely due to evolution in these processes, and not the increasing or decreasing strength of the background with redshift.

\begin{figure}
\includegraphics[clip,trim={0.0cm 0.0cm 0.0cm 0.0cm}, width=0.99\linewidth]{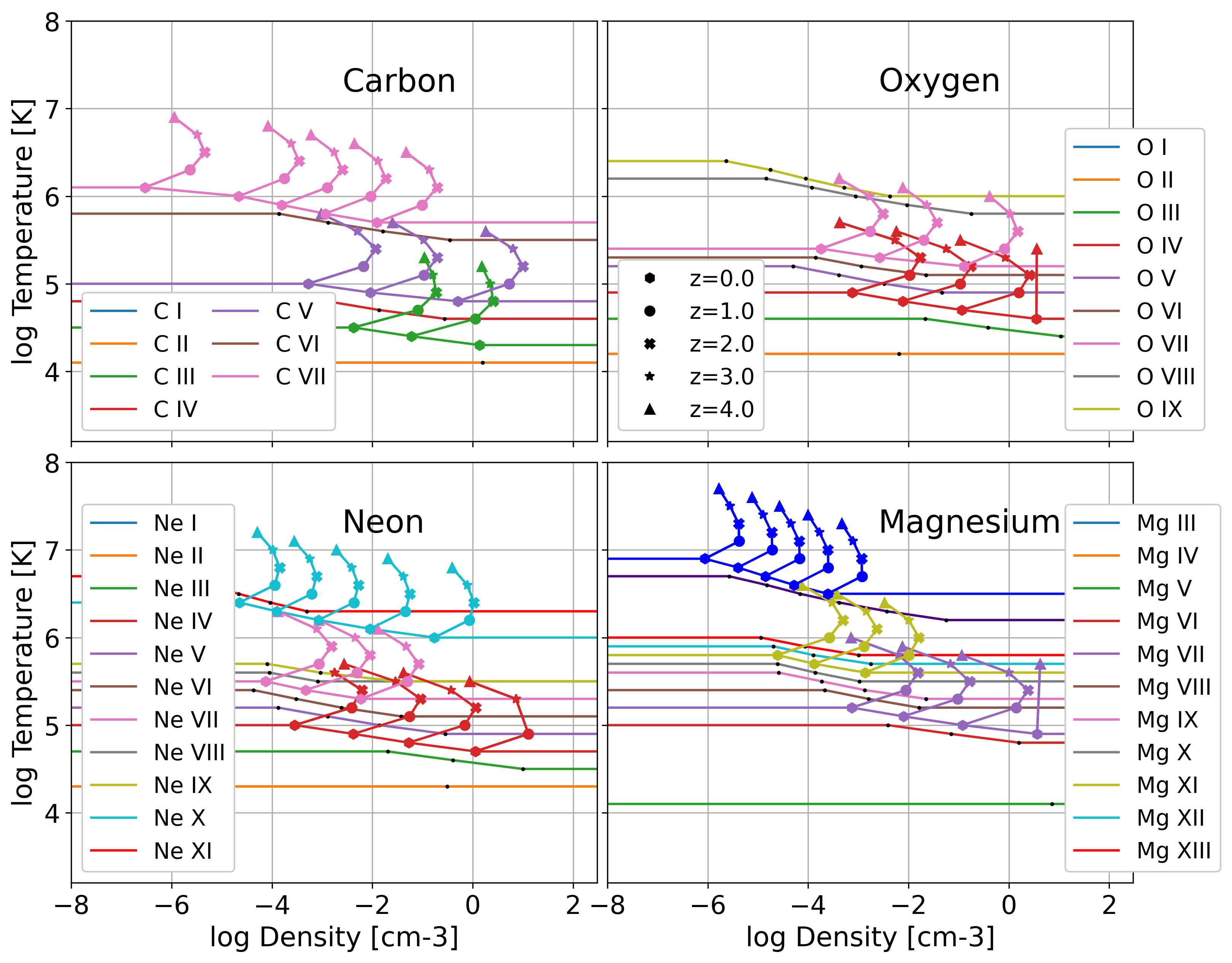}\par 
\caption{The change with redshift in the transition densities for selected ions. The vertical offset is just for visibility, but the temperature of the transitionary points stays fixed with redshift. Different symbols show different redshifts.} 
\label{fig:dens-zdep}
\end{figure}

The density dependence on the ionizing background is somewhat more noticeable, but only under the constraint that the cutoffs were already only weakly dependent on density (i.e., they were density dependent only within a small range of transitionary temperatures). In Fig. \ref{fig:dens-zdep} we show the effect of redshift on the density thresholds for the transitionary points. Tracing from $z=0$ to $z=4$ we see that the density threshold for each ion drifts to the right from $z=0$ to $z=2$, reflecting the increasing strength of the ionizing background at that redshift, and then drifts back to the left from $z=2$ to $z=4$, as the ionizing background decreases. A few species have their transition temperatures changed, and so do not appear at all 5 redshifts. Vertical offsets are added for increased visibility. 

Note that the shapes are slightly different between different ions, so wavelength-specific effects of the changes from $z=4$ to $z=0$ can cause minor changes to their relative abundances, but the main effect of increased background activity at $z\sim2$ is just decreasing the effective density of the gas for the purposes of ion fraction calculations. This change has noticeable effects on ion fraction when an ion is PI, and basically has no effect if the ion is CI.

\section{Ionization Mechanism cutoffs as a function of ionization Energy}\label{sec:ionization-energy}

The very limited effects of redshift and ionizing background on key ionization mechanism suggests that there is a physical reason for the cutoffs being where they are which does not depend strongly on the ionizing background. In this section we show that the PI-CI cutoff for an ion appears to be determined primarily by the ion's ionization energy. 

\begin{figure}
\includegraphics[clip,trim={0.0cm 0.0cm 0.0cm 0.0cm}, width=0.99\linewidth]{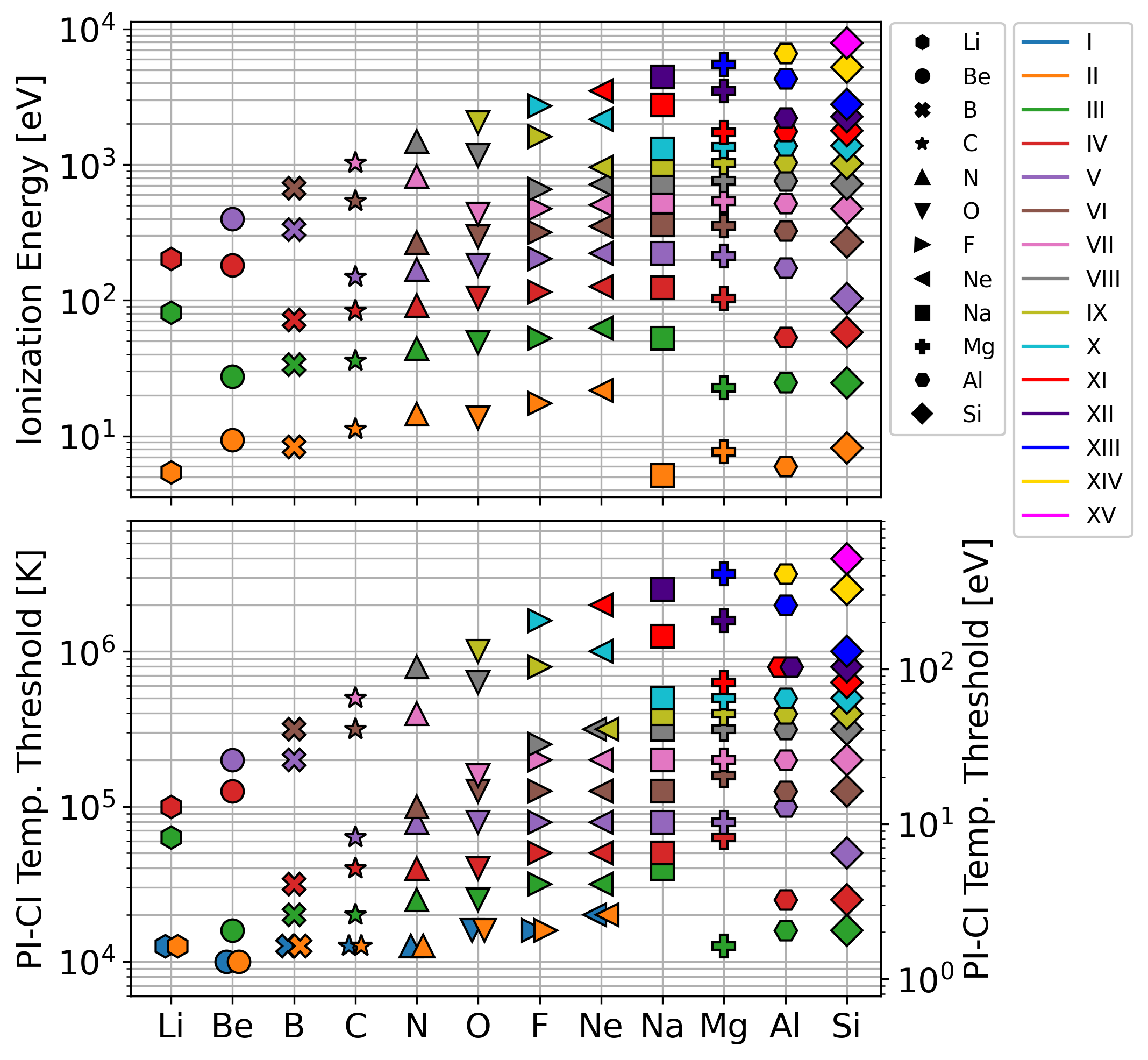}\par 
\caption{Top: The ionization energies for all ions of the first 12 metals. Bottom: The PI-CI cutoff temperatures for the same ions. Different ionization levels are shown with different colors, different elements with different symbols. Note that the PI-CI cutoff is defined for neutral atoms, but ionization energy is not. Atoms which have no transition (see Appendix \ref{sec:appendix}) are not shown.} 
\label{fig:energy-and-cutoffs}
\end{figure}

In the top panel of Fig. \ref{fig:energy-and-cutoffs} we show the ionization energies of all ionization species for the first 12 metals, including those with the highest abundances in the Universe generally and in the CGM. The quantum mechanical ionization structure gives rise to multiple shells of increasing ionization energy from the outside in, and clear gaps form between the energy required to leave an outer shell and the energy required to leave the next innermost shell. In the bottom panel of Fig. \ref{fig:energy-and-cutoffs}, we show the CI-PI cutoff for each ion. Pictured is the minimum temperature level for a transition point, that is the first temperature at which CI gas is a majority at any density. Changing the threshold to represent the CIE peak or the first temperature which is CI at all densities has only minor effects on this conclusion. The left axis is in temperature units (K) and the right axis in units of energy per particle (eV), following $\left<E\right> = \frac{3}{2}k_b T$. We see here that the same shell structure is replicated in both.

\begin{figure}
\includegraphics[clip,trim={0.0cm 0.0cm 0.0cm 0.0cm}, width=0.99\linewidth]{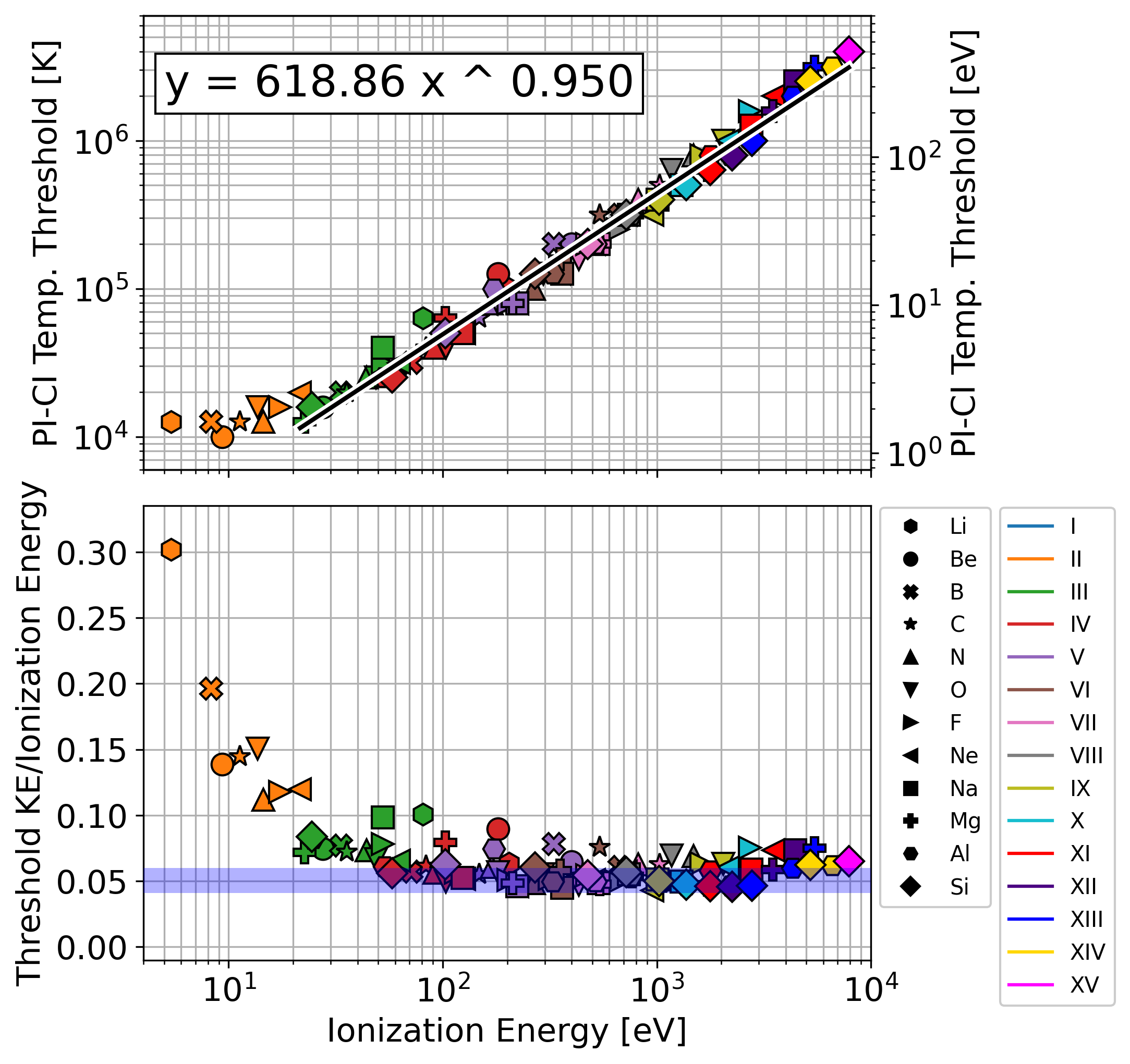}\par 
\caption{Top: The lowest PI-CI cutoff temperature versus ionization energy for all ions of the first 12 metals, except for those without transitions (Appendix \ref{sec:appendix}). The boxed equation is the best-fitting power law (black line). Bottom: The PI-CI cutoff temperatures divided by ionization energy for the same ions. Shaded blue region shows prediction from equation (\ref{eq:num_est}). It is clear that ionization mechanism linearly depends on ionization energy.}
\label{fig:energy-over-cutoffs}
\end{figure}

In Fig. \ref{fig:energy-over-cutoffs} we compare the two numbers directly, and see that they have an almost perfectly linear relationship. In the top panel, we see that the best-fitting line in log-log space has a slope of 0.950. We fit the line to all states except the singly ionized states, which appear to have a slightly different relationship, which might be due to the fact that the singly-ionized state has a threshold which is somewhat more challenging to define than the further states, as described in Section \ref{sec:neutral-single-ionized}. The linear relationship suggests that the meaningful quantity of interest to determine whether an ion is PI or CI is the ratio between the average energy per particle and the ionization energy. We see in the bottom panel of Fig. \ref{fig:energy-over-cutoffs} that over almost three orders of magnitude in ionization energy, the threshold energy for being CI-dominated is consistently around 4 -- 7 percent of the average energy per particle, increasing to up to $\sim 30$ percent for the singly ionized states. For higher ions, this result aligns quite well with the predicted value (blue), derived in Section \ref{sec:ionization-derivation}.
\subsection{Derivation of relationship between CI threshold and ionization energy}
\label{sec:ionization-derivation}
The law shown above in Fig. \ref{fig:energy-over-cutoffs} can be approximately derived as a consequence of, above all, the steeply exponential dependence of ion fraction on temperature in CIE. In ionization equilibrium, the ionization rate of each species is set equal to the recombination rate for the species ionized one additional time. Generally, there are many mechanisms governing both of these rates, including photoionization, collisional ionization, radiative recombination, collisional recombination, and what are called `charge-exchange reactions' which change the ionization of two species simultaneously. We are interested in the PI-CI cutoff, which takes place at the lowest temperature in which CI fractions are relevant. In this regime, the density is taken to be high enough that the metagalactic background is negligible, and there are no significant local sources of ionization, so the photoionization term is neglected. However since even the highest densities in the CGM are much less than those studied in the ISM or solar system environment, we can also neglect the contributions of collisional recombination and charge-exchange interactions \citep{ house_ionization_1964,dopita_astrophysics_2003}. This density regime is known as the `coronal approximation', where ion ratios are determined by the following equation:
\begin{ceqn}
\begin{align}
    \label{eq:coronal_approx}
    \frac{n_{j+1}}{n_j} &= \frac{C_{j,j+1}}{\alpha_{j+1,j}} \ ,
\end{align}
\end{ceqn}
where $n_j$ is the concentration of ions with $j$ electrons removed from the neutral state, $C_{j,j+1}$ is the collisional ionization rate from state $j$ to state $j+1$, and $\alpha_{j+1,j}$ is the radiative recombination rate from the higher state to the lower one. 

Following \citet{house_ionization_1964}, we will use the rates given by \citet{allen_solar_1961} and \citet{elwert_verallgemeinerte_1952}.

\begin{ceqn}
\begin{align}
    \label{eq:CI_rate}
    C_{j,j+1} &= 2.47\cdot 10^{-8} A \zeta_j n_e \left(\frac{kT}{\varepsilon_H}\right)^\frac{1}{2}\left(\frac{\varepsilon_H}{\chi_j}\right)^2e^{\frac{-\chi_j}{kT}}
\end{align}
\end{ceqn}
where
\begin{ceqn}
\begin{align*}
    A &= 3.1-\frac{1.2}{Z_j}-\frac{0.9}{Z_j^2}\\
    \zeta_j &= \textrm{number of electrons in outer shell,}\\
    n_e &= \textrm{unbound electron density,}\\
    \chi_j &= \textrm{difference in ionization energies, } \varepsilon_{j+1} \textrm{ and } \varepsilon_j\\
    Z_j &= \textrm{ionic charge after ionization,}\\
    \varepsilon_H &= \textrm{ionization energy of hydrogen,}
\end{align*}
\end{ceqn}
and 
\begin{ceqn}
\begin{align}
    \alpha_{j+1,j} &= 5.16\cdot 10^{-14} f_1 n g n_e \left(\frac{\varepsilon_H}{kT}\right)^\frac{1}{2}\left(\frac{\chi_j^2}{\varepsilon_H kT}\right)e^{\frac{\chi_j}{kT}}E_1 \left(\frac{\chi_j}{kT}\right),
\end{align}
\end{ceqn}
where 
\begin{ceqn}
\begin{align*}
    n &= \textrm{quantum number of ground state,}\\
    E_1(x) &= \int_x^\infty \frac{e^{-t}dt}{t} \textrm{ is the first exponential integral.}
\end{align*}
\end{ceqn}
$f_1$ and $g$ are empirical factors of $O(1)$. The ion ratio is thus 
\begin{ceqn}
\begin{align}
    \label{eq:ionratio}
    \frac{n_{j+1}}{n_j} = B \frac{\left(\varepsilon_H kT\right)^2}{\chi_j^4}\frac{e^{\frac{-2\chi_j}{kT}}}{E_1 \left(\frac{\chi_j}{kT}\right)} \ ,
\end{align}
\end{ceqn}
where the unitless value $B = 4.79 \cdot 10^5 \left(\frac{A\zeta_j}{f_1 n g}\right)$ is constant with respect to $T$ and $n_e$, and only mildly varies with species. We will treat it as a constant of $O(10^6)$ in this rough analytic calculation. 

Successively applying equation (\ref{eq:ionratio}) can give $n_j$ for each of an ion's $k$ possible states in terms of $n_0$. Ion fractions can be found with

\begin{figure}
\includegraphics[clip,trim={0.0cm 0.0cm 0.0cm 0.0cm}, width=0.99\linewidth]{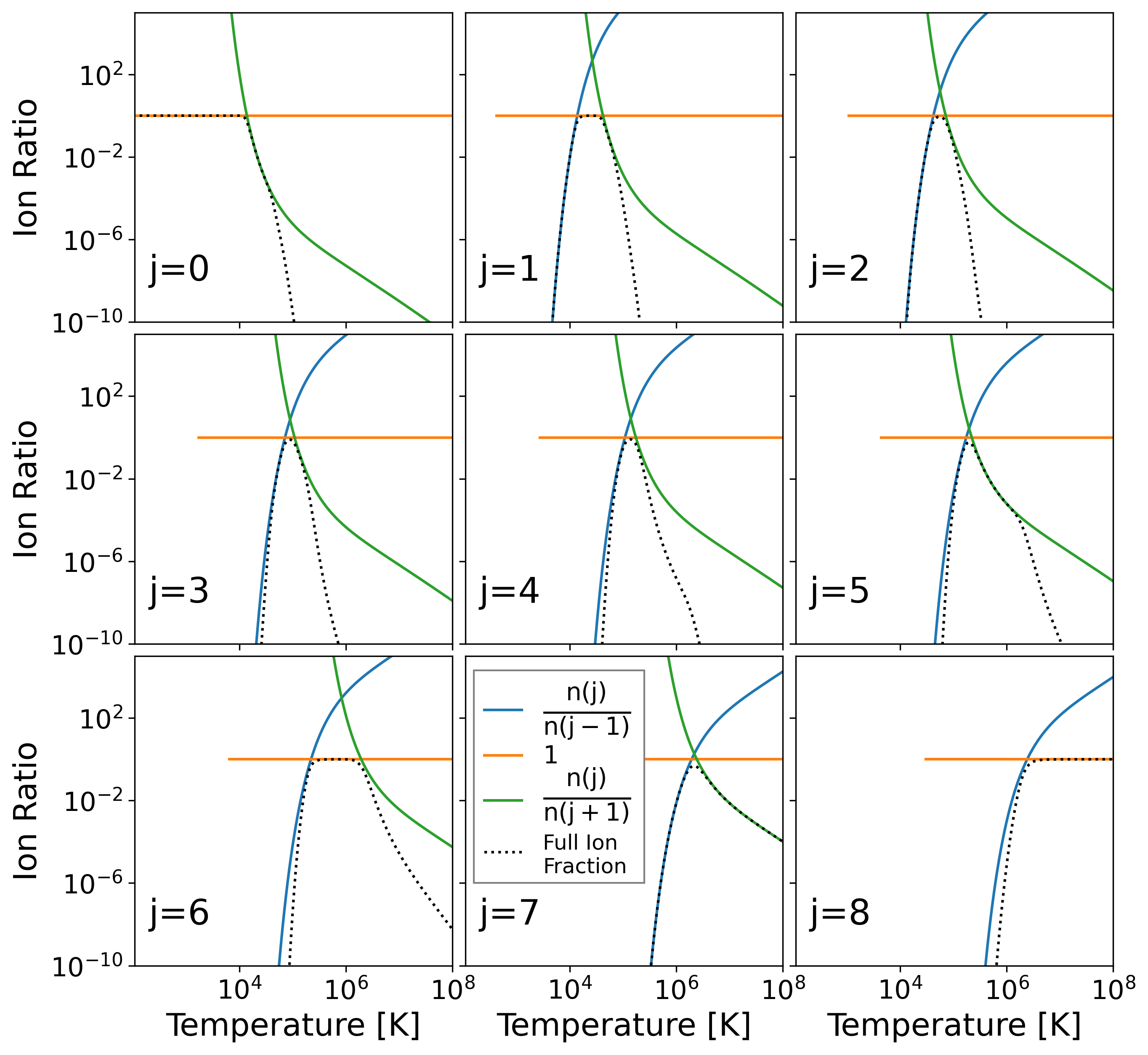} 
\caption{Demonstration of the three-term approximation used in equation (\ref{eq:three_term_approx}) for all species of oxygen. We see here $n_j$ vs $n_{j-1}$ and $n_{j+1}$, written out with parentheses, e.g. $n(j)$, for legibility. These ratios are compared to the actual ion fraction when using all nine terms (dotted line). Each term is calculated using equation \ref{eq:ionratio} with a constant $B=10^6$.} 
\label{fig:coronal_approx}
\end{figure}

\begin{ceqn}
\begin{align}
    f_j &= \frac{n_j}{n_0+n_1+\dots+n_j+\dots+n_k}\nonumber \\
    &= \frac{1}{\frac{n_0}{n_j}+\frac{n_1}{n_j}+\dots+1+\dots+\frac{n_k}{n_j}}\nonumber\\
    &\approx\frac{1}{\frac{n_{j-1}}{n_j}+1+\frac{n_{j+1}}{n_j}} \ ,
    \label{eq:three_term_approx}
\end{align}
\end{ceqn}
where the last line approximates the denominator in the neighborhood of the CIE peak to be dominated by the $j-1$, $j$, and $j+1$ terms, which can be considered as governing the behavior where the fraction is rising, flat, and falling with increasing temperature (see the right-hand side of Fig. \ref{fig:contours}). In Fig. \ref{fig:coronal_approx}, we show each of the ratios used in this approximation separately, compared to the full ion fraction including all terms. This shows the ratio $\frac{n_j}{n_{j-1}}$ term accurately tracks the rising ion fraction at temperatures below the peak, while the ratio $\frac{n_j}{n_{j+1}}$ term tracks, somewhat less effectively, the falling ion fraction at temperatures above the peak. 

Combining equations (\ref{eq:ionratio}) and (\ref{eq:three_term_approx}), and taking the low-T approximation so that $E_1 \left(\frac{\chi_j}{kT}\right)\approx e^{\frac{-\chi_j}{kT}}$ and the left (rising) term of the denominator dominates, this equation simplifies to
\begin{ceqn}
\begin{align}
    f_j &\approx \frac{n_j}{n_{j-1}} \\
    &\approx C\left(j,T\right)e^{\frac{-\varepsilon_j}{k_B T}},
    \label{eq:finalapprox}
\end{align}
\end{ceqn}
where $C\left(j,T\right)=\frac{B\left(\varepsilon_HkT\right)^2}{\chi_{j-1}^4}e^{\frac{\varepsilon_{j-1}}{kT}}$ encapsulates all dependence besides the (larger) exponential. We are interested in the first temperature for which this fraction is non-negligible, which takes place in our algorithm at around $f_j=10^{-8}$. Even if $B$ is taken as constant, this equation is clearly transcendental and depends on both $T$ and $\varepsilon_{j-1}$, and therefore cannot be solved analytically. However, the dependence of the fraction on $C\left(j,T\right)$ is weak enough that it is sufficient to note its order of magnitude in the relevant temperature region, which is $C\left(j,T\right)=O(10^3-10^8)$.

Then, we see that $f_j$ first passes $10^{-8}$ at roughly
\begin{ceqn}
\begin{align}
    \frac{3}{2}kT &=\frac{0.65}{8+\log\left[C\left(j,T\right)\right]}\varepsilon_j \nonumber\\
    &= \left(0.041 - 0.059\right)\varepsilon_j,
    \label{eq:num_est}
\end{align}
\end{ceqn}
where the low end and the high end of this range involve taking $C\left(j,T\right) = 10^8$ and $C\left(j,T\right) = 10^3$, respectively. This prediction aligns remarkably well with the detected trend shown in Fig. \ref{fig:energy-over-cutoffs}, which had cutoffs that varied between 4 percent and 7 percent of $\varepsilon_j$.

\section{Consequences for CGM Modeling and Interpretation}
\label{sec:split-model}

We will now analyse whether our physically motivated distinction between PI and CI ions in a single gas parcel can give a meaningfully good approximation for studies of the CGM. Essentially, older studies generally assume gas is in either a PIE state \citep[e.g.][]{stern_universal_2016}, or a CIE state \citep[e.g.][]{faerman_massive_2017,faerman_massive_2020}. The power of making this approximation is obvious: if a component with multiple ions is a single phase in PIE, then the density can be constrained not only by the equivalent widths, which depend on sensitive geometric constraints and noisy spectral resolution, but also by the ratios between different ions, which can be more robust. On the other hand, if this component is in CIE, then instead the temperature can be constrained in an analogous way, increasing the power of the line shape and structure to determine the density independently. In recent years, more researchers are becoming aware that both PI and CI mechanisms can make meaningful contributions for almost every ion, depending on the gas phase it is found in. While it is possible to attempt to constrain both the temperature and the density via Voight profile $b$ parameters and equivalent widths, it is very noisy and hard to sort accurately into phases. A more modern approach, which might be used more commonly in the future, is that of \cite{haislmaier_cos_2021}. They did not assume either CIE or PIE, but instead allowed multiple phases to exist in the same components, using one or both mechanisms, and used Monte Carlo simulation techniques to identify the best-fitting density and temperature for each phase. However, this addition of multiple phases leads to possibly unnecessarily increased complexity of the final state. 

\begin{figure*}
\includegraphics[clip,trim={0.0cm 0.0cm 0.0cm 0.0cm}, width=0.99\linewidth]{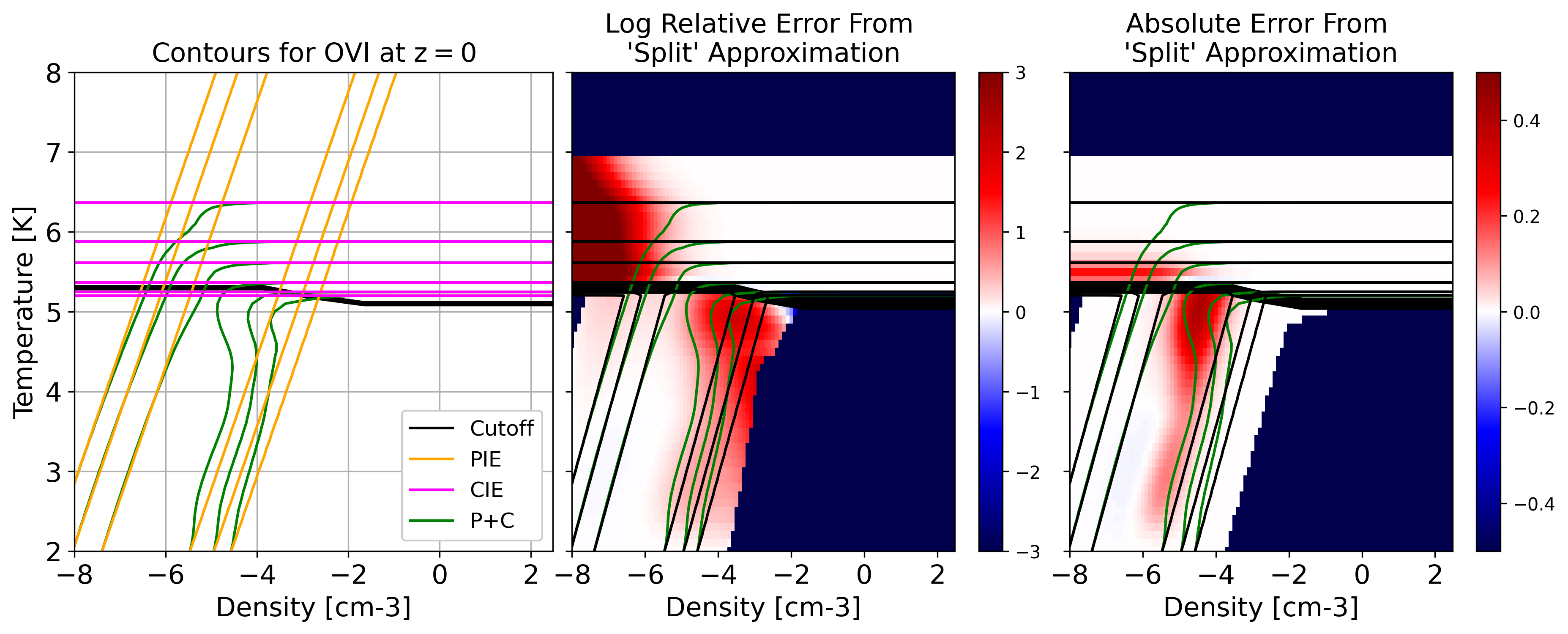}\par 
\caption{Left: Comparison of a pure PIE, pure CIE, and combined (`P+C') output for O~{\sc vi}. The `split' approximation is defined to be PIE below the cutoff, CIE above. Middle: comparison of `split' (black lines) approximation to {\sc Cloudy} ion fraction table. This panel shows the logarithm of $f_{\rm split}/f_{\rm P+C}$, so red indicates an overestimation, blue an underestimation, and white approximately correct. Dark blue, however, represents regions which are negligible in both approximations. Right: same as middle, but showing the absolute difference $f_{\rm split} - f_{\rm P+C}$. This figure shows our approximation is significantly better than naive assumption of one or the other mechanism, although errors remain at the low-density CIE end and high-density PIE end.
} 
\label{fig:split-approx}
\end{figure*}

Using a strict definition of PI vs CI gas can allow the power of the old approach, which used ion ratios to get relatively clean estimates of density and temperature from PIE and CIE, respectively, to be incorporated into a system where clearly both ionization mechanisms matter. A suggested workflow for analysis of complex spectra with multiple ions is as follows. By getting a rough estimate of the temperature, or even guessing a temperature and iterating over multiple guesses, all ions can be assigned PI or CI prior to fitting. Then, if the detected ions are PI, they can be fit to a particular density, and if they are CI, they can be fit to a particular temperature. Ions which are transitionary at this rough temperature can be ignored on this first pass, to be fit later for additional precision. 

In Fig. \ref{fig:split-approx} we show the errors involved in this approach (and by extension, involved in prior studies using pure PIE or pure CIE approaches). Here we compare the actual distribution, from iterating over a grid of simulations run through {\sc Cloudy}, to the approximation we will call the `split' distribution. This distribution is created by assuming density-independent CIE above the defining line, and (nearly) temperature-independent PIE below the line. PIE is never fully temperature independent, rather it is assumed that the contours follow straight powerlaws in log-log space with a slope of $\gamma = \frac{5}{3}$. This dependence springs from the fact that the effective absorption resonance gets larger with higher temperature, as the increased Doppler broadening not only increases a species' receptiveness to absorption lines, but also to ionizing radiation. This dependence is quite consistent and is generally included in PIE modeling by using the ionization parameter $U$.

On the left panel, we simply show these two distributions, the simulated and the approximate one, and in the middle and right panels we show the relative and the absolute difference, respectively. Red pixels on each graph indicate where the `split' distribution overestimates the ion fraction compared to the simulation, and blue pixels show where it instead underestimates the real distribution. Dark blue pixels represent where both approximations give negligible values ($f<10^{-6}$), so the difference is not meaningful.

There are two major error-prone regions to be careful of with this approach. The first is somewhat obvious. CIE is density independent, but as we saw in Fig. \ref{fig:fractions}, every CI ion except for fully ionized states collapses in fraction at low enough density, at all temperatures. Every ion drops off as photoionization destroys their numbers at low density, and our definition did not give a role to PI there because it only destroys, and does not create, that ion. So, the CI region on the left end of the graph vastly overestimates each ion. The second is the PI region on the right hand side of the PIE peak. Errors here are a result of the fact that we have defined the density cutoff for transitionary gas to specifically refer to the point where, at a given density, 50 percent of ions are created through each mechanism. Clearly this approximation will lead to notable errors near this point, as it effectively asserts that on the left of the 50 percent mark, 100 percent of the ions are created through PI, and on the right it is 0 percent. Similarly, right below the first transitionary temperature are usually several temperatures where some ions are created through PI, but do not quite reach 50 percent, which is also approximated as 0 percent PI.

This approximation of a CIE and PIE `split' thus functions most effectively in the low-density PIE limit, and the high-density CIE limit. But even in regions with substantial errors, it remains a better approach than assuming a mechanism which might be totally wrong, as is traditionally done by both observers and modellers. It is also important to note that there are only small regions in which high and low ions are likely to both coexist and be created through different mechanisms, and that is precisely where the low-density CIE regime, for the low ion, overlaps the high-density PIE regime, for the high ion. Thus, unfortunately, this definition is most relevant exactly where the errors are highest. 

\section{Summary and Conclusions}\label{sec:conclusion}

In this work we refined a novel definition of PI and CI gas, which has previously been shown in early forms in \cite{roca-fabrega_cgm_2019} and \cite{strawn_o_2021}. This definition allows ions to be identified individually, without assuming a universal ionizing mechanism -- i.e., PIE or CIE -- but also without foregoing the lessons learned from those two regimes and demanding use of an arbitrary 2D ion fraction table. Analyzing each ion at fixed temperature, we define an ion at that temperature to be PI if its fraction-density curve has a maximum and does not stabilize at high density, CI if it stabilizes at high density and does not have a maximum, and transitionary if it has both a maximum and a high-density shelf. In the transitionary case, the majority of the ion is created through CI at high density, and PI at low density. Further examination of the consequences of this definition using {\sc Cloudy} showed several key insights, which are often ignored in existing CGM literature. 

\ 

\noindent The main results are as follows:
\begin{itemize}
\item {\bf Temperature Threshold:} Most ions have only a few or no transitional temperatures, covering less than 1 order of magnitude in temperature. A good first approximation, then, is that the difference between PI-dominated and CI-dominated gas is a temperature threshold. Higher ions have a transition at a higher temperature, and thus in a single cloud at moderate temperature, high ions can be created through PI and low ions through CI, which is the reverse of what is often assumed.
\item{\bf Redshift Independence:} Where each mechanism dominates in phase space has only a slight dependence on redshift from $z=0$ to $z=4$, and on ionizing background radiation more generally, including little change with variation of the slope $\alpha_{UV}$ to include both a much harder and softer potential background. Even though the \textit{fractions} induced by PI change with redshift, \textit{which} mechanism dominates in a certain region barely changes in temperature at all, and only changes moderately in density at the few transitionary temperatures.
\item{\bf Ionization Energy Relation:} The temperature cutoff for CI dominance is strongly determined by ionization energy. Nearly all ions become CI-dominant when the average kinetic energy per particle exceeds roughly six percent of the ionization energy. This result is a straightforward consequence of the steep temperature dependence of the coronal approximation used in CIE. The exception to this trend is singly-ionized states, which require a larger fraction.
\item{\bf `Split' distribution:} An approximation which assumes ions are distributed into roughly their CIE fractions above the cutoff temperature, and roughly their PIE fractions below it, is an adequate approximation of the full, complex distribution, especially at the high-density CIE end and the low-density PIE end. 
\end{itemize}

Future applications of this framework can be used for help interpreting complex spectra such as the CASBaH survey \citep{CASBaH}. This depiction could, for instance, radically improve the priors used for phase assignment as in \cite{haislmaier_cos_2021}, which found both PIE and CIE necessary to replicate CASBaH absorption components.

\section*{Acknowledgements}
Partial support for CS was provided by grant HST-AR-14578 to JP from the STScI under NASA contract NAS5-26555 and from JP's Google Faculty Research Grant. CS also received support from the UCSC Science Internship Program (SIP) as well as the ARCS Foundation. SRF acknowledges support from a Spanish postdoctoral fellowship, under grant number 2017- T2/TIC-5592. SRF also acknowledges financial support from the Spanish Ministry of Economy and Competitiveness (MINECO) under grant number AYA2016-75808-R, AYA2017-90589-REDT and S2018/NMT-429, and from the CAM-UCM under grant number PR65/19-22462. We would like to thank Christophe Morriset for many helpful discussions of {\sc Cloudy} and the various mechanisms which it uses, and would also like to acknowledge the many other developers of that software. We also benefited from helpful discussions with J. Xavier Prochaska, Joe Burchett, Sandra Faber, David Koo, Anatoly Klypin, Joanna Woo, and the SIP interns working with CS, Paul Mayerhofer, Soumily Maji, and Antonio Man. Finally, we thank an anonymous referee who helped us clarify several key points of the paper.

\section*{Data Availability}
Data underlying this article is available at \url{https://github.com/claytonstrawn/pi_or_ci}. This site includes a notebook for directly generating all images used in this paper, as well as scripts to apply this definition to other studies.

\bibliographystyle{mnras}
\bibliography{strawn22bib.bib}

\appendix
\section{Low Ions of Heavy Elements}
\label{sec:appendix}
In Sections \ref{sec:definition}, \ref{sec:redshift-dependence}, and \ref{sec:ionization-energy}, the neutral and singly ionized species of Mg, Na, Al, and Si were generally not shown. This omission is because at the low temperatures which would presumably house the transition points for these ions, they are not adequately characterized by the shapes described in Section \ref{sec:definition}. While some of them appear to follow similar patterns, and indeed the algorithm as initially written does classify them as one or the other, they do not have several of the features which should be present in either PI or CI, and the temperatures of their `cutoffs' were far out of line with all the predictions in Section \ref{sec:ionization-derivation}.

\begin{figure*}
\includegraphics[clip,trim={0.0cm 0.0cm 0.0cm 0.0cm}, width=0.99\linewidth]{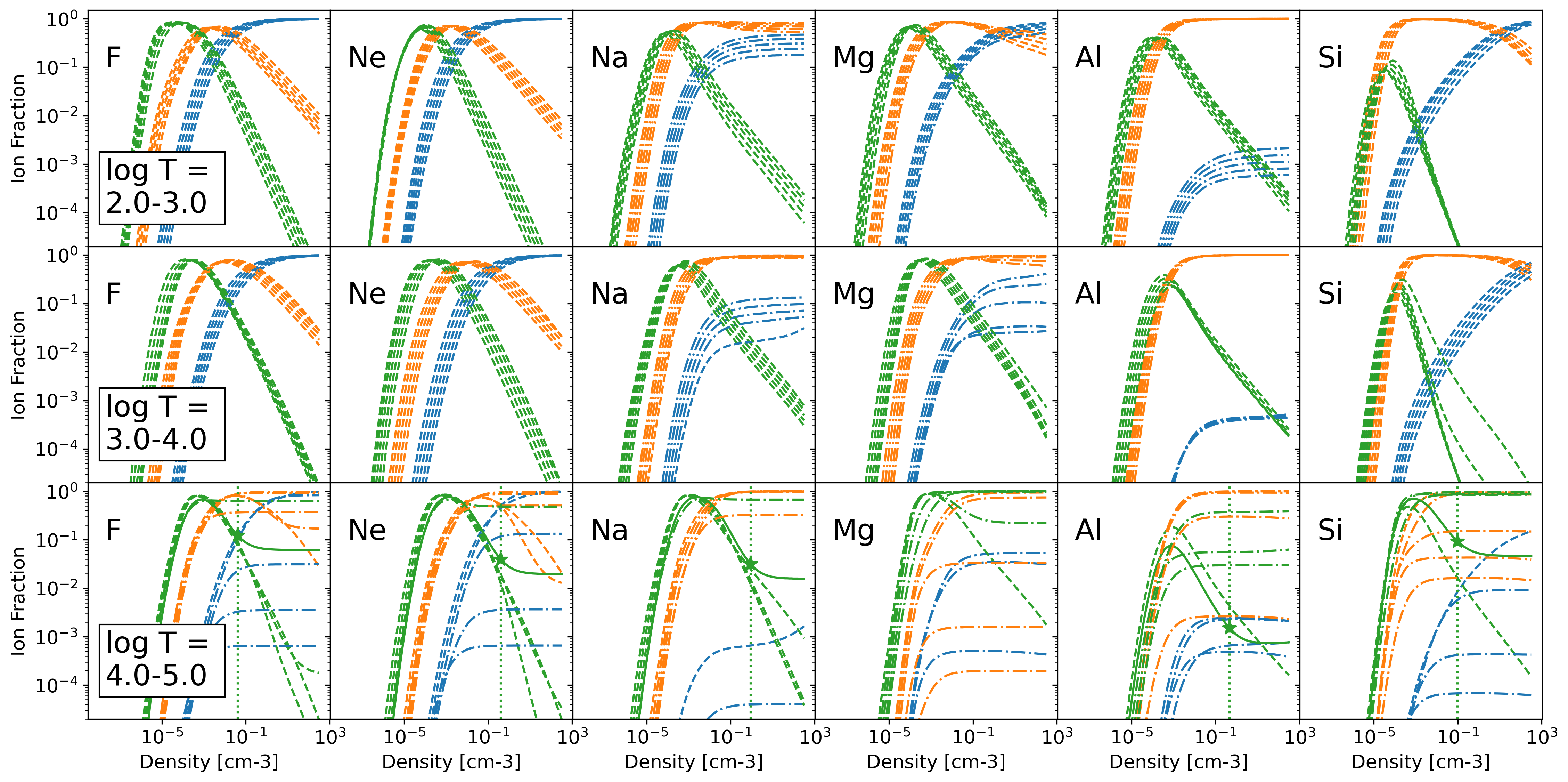}\par
\caption{Like Fig. \ref{fig:fractions}, but showing neutral (blue), singly-ionized (orange), and doubly-ionized (green) species of the six largest atoms studied here. Linestyle represents the naive sorting of each ion into PI (dashed), CI (dot-dashed), or transitionary (solid). In each cell are 5 lines of each color, representing increments of 0.2 dex in temperature.}
\label{fig:specialcase}
\end{figure*}

To explore the new processes that appear, in Fig. \ref{fig:specialcase} we analyse a large portion of the ion fraction grid, showing fraction with density at a wide range of temperatures. The leftmost two columns show F and Ne, which we will consider `light elements' because they are in the second row of the periodic table. The rightmost four columns show Na, Mg, Al, and Si, which are `heavy elements', in the third row. In each panel, five lines are shown for each at increments of 0.2 dex in temperature. While they are not labeled individually, they do follow some expected trends, (i.e. neutral fractions always decrease with increasing temperature, though see point (iii) below).

Essentially, the usual case (light elements) is naive PIE at $T<10^4$ K. At high enough densities, ion fraction for the neutral state approaches 1, all other states approach 0. Each state has a peak at some characteristic density, and while the peaks are not always exactly the same height or width, each ion is dominant around its own peak, with at least 50 percent of the total. At higher temperatures, ions transition to CIE on the high-density side, with characteristic flat shelves even for low ions.

In contrast, there are several strange behaviors for the heavy elements which do not appear to follow the `universal' patterns.
\begin{enumerate}
    \item Na~{\sc i} and Al~{\sc i} are never dominant over Na~{\sc ii} and Al~{\sc ii}, even at the highest densities and lowest temperatures. This outcome is of course possible with a CI classification, however Na can be seen to actually curve upwards at high densities at around $10^4$ K, showing this `CI' region is not at all density-independent. Al~{\sc i} does the same when extended to higher densities, though not shown here. 
    \item Mg~{\sc ii} and especially Si~{\sc ii} have extremely wide peaks, even reaching a long plateau before declining slightly at very high densities. This plateau is effectively density independent, and therefore the high-density decline would is not simply according to the decreasing strength of the ionizing background as a `PI' classification would assume, but due to the trade-off between the photoionization mechanism and another mechanism.
    \item Mg~{\sc i} and Al~{\sc i} have large regions in temperature space where, while the shape appears `CI', changes in temperature have no effect on the fraction. The 7 lines for Al~{\sc i} between $10^3$ and $10^{4.5}$ K are all overlapping, and Mg~{\sc i} fractions from $10^{3.5}$ and $10^{4.4}$ K have very little movement.
\end{enumerate}

All these effects take place because the assumption that ions can be categorized as a binary of `primarily PI' and `primarily CI' relies on the fact that the no other mechanism is relevant, even though there are a variety of both ionization and recombination processes studied in the literature and implemented in {\sc Cloudy}. In densities relevant to the CGM, usually the only relevant ionization mechanisms are photo and collisional ionization, and the only relevant recombination process is radiative recombination. Radiative recombination cancels out the density dependence of collisional ionization rates, leaving `CI' ions completely density independent at fixed temperature, while it does not cancel for `PI' ions, giving rise to simplified peaks at fixed temperature. 

The other ionization processes relevant in astronomy include the Auger process and charge transfer, while other recombination processes include dielectronic processes, three-body recombination, and charge transfer \citep{ferland_cloudy_1998,dopita_astrophysics_2003,kallman_photoionization_2021}. For the lowest ionization states of heavy elements, specifically those with valence electrons in the 3n shell, electrons are not tightly enough bound to the nucleus to effectively resist these other processes. This susceptibility is not only because they have low ionization energies, but also because their electrons have larger average distance to the nucleus, and lower average speed, and so are easier to interact with. A detailed study of this regime, including analysis of whether this regime is relevant in the CGM at all, will almost certainly be much more complicated than the PI and CI binary explored here, and is left for future work. 

The main use case of the approximation presented in Section \ref{sec:split-model} remains for ions which are ionized more than once, including fully-ionized states. To some extent (Fig. \ref{fig:energy-over-cutoffs}) even the lighter elements have difficulty following the trends for neutral and singly ionized ions, however the definition at least is coherent and consistent. For these heavy elements, it is neither, and they effectively show the limits of where this approximation is appropriate.







\bsp	
\label{lastpage}

\end{document}